\documentclass{amsart}

\usepackage{graphicx,epsfig}

\usepackage{amsmath,amssymb,latexsym, amsfonts, amscd, amsthm, xy}

\usepackage{url}

\usepackage{hyperref}

\usepackage{mathrsfs}

\input{xy}
\xyoption{all}

\hypersetup{linkcolor=red}

\hypersetup{linkcolor=blue}

\makeindex 

=630pt \headheight = 12pt \headsep = 20pt

\hypersetup{
    colorlinks   = true,
      citecolor    = red
}

\hypersetup{linkcolor=red}

\hypersetup{linkcolor=blue}

\theoremstyle{plain}
\newtheorem{theorem}{Theorem}[section]

\theoremstyle{definition}
\newtheorem{definition}[theorem]{Definition}
\newtheorem{remark}[theorem]{Remark}
\newtheorem{example}[theorem]{Example}

\def\bR{\mathbb{R}}

\def\fb{\mathfrak{b}}

\def\fa{\mathfrak{a}}

\def\fe{\mathfrak{e}}

\begin{document}
\definecolor{myblue}{RGB}{80,80,160}
\definecolor{mygreen}{RGB}{80,160,80}

\title{Weight Prediction for Variants of Weighted Directed Networks}

\author{Dong Quan Ngoc Nguyen, Lin Xing, and Lizhen Lin}

\date{September 29, 2020}

\address{Department of Applied and Computational Mathematics and Statistics \\
         University of Notre Dame \\
         Notre Dame, Indiana 46556, USA }

\email{\href{mailto:dongquan.ngoc.nguyen@nd.edu}{\tt dongquan.ngoc.nguyen@nd.edu}}

\urladdr{http://nd.edu/~dnguye15}

\email{\href{mailto:lxing@nd.edu}{\tt lxing@nd.edu}}

\email{\href{mailto:lizhen.lin@nd.edu}{\tt lizhen.lin@nd.edu}}
\maketitle

\begin{abstract}
A weighted directed network (WDN) is a directed graph in which each edge is associated to a unique value called weight. These networks are very suitable for modeling real-world social networks in which there is an assessment of one vertex toward other vertices. One of the main problems studied in this paper is prediction of edge weights in such networks. 
We introduce, for the first time, a metric geometry approach to studying edge weight prediction in WDNs. We modify a usual notion of WDNs, and introduce a new type of WDNs which we coin the term \textit{almost-weighted directed networks} (AWDNs).  AWDNs can capture the weight information of a network from a given training set.  We then construct a class of metrics (or distances)  for AWDNs which equips such networks with a metric space structure. Using the metric geometry structure of AWDNs, we propose modified $k$ nearest neighbors (kNN) methods and modified support-vector machine (SVM) methods which will then be used to predict edge weights in AWDNs.   In many real-world datasets, in addition to edge weights, one can also associate weights to vertices which capture information of vertices; association of weights to vertices especially plays an important role in graph embedding problems. Adopting a similar approach, we introduce two new types of directed networks in which weights are associated to either a subset of origin vertices or a subset of terminal vertices . We, for the first time, construct novel classes of metrics on such networks, and based on these new metrics propose modified $k$NN and SVM methods for predicting weights of origins and terminals in these networks. We provide experimental results on several real-world datasets, using our geometric methodologies. 
\end{abstract}

\section{Introduction}
Many real world datasets can be modeled as \textit{weighted directed networks} (WDNs) which are the main objects studied throughout this paper. In a WDN, each edge is associated to a weight (which is a real number between a given closed interval $[a, b]$ in $\bR$.). For example, in a digraph $G = (O, T, E)$, where $O$ is the set of buyers, and $T$ is the set of products, an edge $\fe = (o, t)$ is formed in $G$ if the buyer $o$ buys the product $t$. A natural question in this social network is \textit{how much} a buyer $o$ likes or dislikes a product $t$. The degree of liking varies, and so it is natural to specify a value to each liking. Hence one obtains a map $W : E \to [a, b]$, where $[a, b]$ is a closed interval in $\bR$ which represents the level of intensity of the evaluation of users towards products.

Many real-world datasets, for example, Bitcoin exchanges or Wikipedia networks are explicit WDNs. A natural question in social network analysis is how to predict weights of edges in WDNs. Edge weight prediction plays an important role in other tasks in  networks such as community detection \cite{TB, YG}, anomaly detection \cite{KSS, KHMKSF}, information diffusion \cite{BRMA, SJ}, among others. Thus a natural question as to how to predict weights of edges in WDNs is important in network analysis. In this paper, we not only deal with the problem of predicting the weight of edges in WDN datasets, but also, for the first time, propose methodology for predicting weights associated with the vertices, i.e., origins and terminals  of the graphs in digraph datasets. Note that prediction of weights of vertices plays an important role in other problems in network analysis, for example, graph embedding problems.  

In order to achieve these aims, we introduce several classes of metrics (or distances) on digraphs which equip such digraphs with structures of metric spaces. Using these geometric structures, we introduce modified $k$ nearest neighbors methods and modified support-vector machine (SVM) methods in such digraphs. Up to the best of our knowledges, this is the the first work that has equipped digraphs with such geometric structures, for the set of edges, the set of origins, or the set of terminals. The geometric approach introduced in this paper may also see applications elsewhere in network analysis \cite{RSH, WAS, LHK}.

The work of \cite{KSSF} focuses on edge weight prediction problem in real-world directed weighted signed networks (DWSNs) which are a special type of WDNs. They introduce two measurements of vertices in DWSNs, the first of which is \textit{fairness}, used to measure how fair a vertex in evaluating other vertices, and the second one is \textit{goodness}, used to measure how good a vertex is from the viewpoints of other vertices. In the real-world DWSN datasets studied in \cite{KSSF}, it is reasonable to view that the weight of an edge as the product of fairness and goodness of two vertices forming the edge. Based on this view, \cite{KSSF} provides an algorithm to predict edge weights in DWSNs. One potential  drawback of this approach is that the edge weight prediction algorithm strongly depends on specific WSN datasets in which the weights of edges are computed using both fairness and goodness values. 

Our geometric approach to edge weight prediction is more direct, and focuses entirely on the weight information of edges instead of depending on nodes features or measurements associated to vertices. In a usual set-up for edge weight prediction, one is given a set of edges, called a \textit{training set}, in a WDN whose weights are explicitly given. A novel feature in our geometric approach is that we incorporate weight information from the training set to introduce a notion of \textit{topological neighborhoods of edges}, and construct a class of metrics in the WDN, using such topological neighborhoods. Many real-world WDNs also carry weights of vertices. For example, in the work of \cite{KSSF}, they introduce two measures of vertices: the \textit{fairness} of an origin vertex captures how fair the origin vertex in assessing other terminal vertices, and the \textit{goodness} of a terminal vertex to signify how good this vertex is assessed by other origin vertices. If we attach fairness and goodness values to the set of origin vertices and terminal vertices, respectively, one obtains two types of directed networks with weighted origins or terminals. These weight information of origins and terminals are important in these real-world social networks; thus a natural question is whether one can also predict weights of origins or terminals in such networks. By adopting a similarly geometric approach, we propose two new types of directed networks which we coin the terms \textit{almost-weighted origin directed networks} (AWODNs) and  \textit{almost-weighted terminal directed networks} (AWWDNs). In AWODNs, a subset of origins with known weights is given. In AWTDNs, a subset of terminals with known weights is given. It bears a resemblance with the notion of AWDNs that we describe above. 

We then propose, for the first time, methodologies using modified kNN and SVM methods for predicting weights of vertices in such networks.  
\section{Variants of Weighted Directed Networks}

In this section, we introduce some basic notions and notation that will be used throughout the paper.

\begin{definition}

A directed graph (or digraph) $G$ is a pair $G = (V, E)$, where $V$ is a set of elements (called vertices of $G$), and $E$ is a collection of ordered pairs of vertices of the form $(u, p)$ (which are called directed edges), where $u, p$ belong in $V$. We indicate the direction of an edge $\fe = (u, p)$ by specifying that $\fe$ starts from the first vertex $u$, and heads to the second vertex $p$. The vertex $u$ is called the \textit{origin} of $\fe$, and the vertex $p$ is called the \textit{terminal} of $\fe$. 

\end{definition}

\begin{figure}
\centering
 \includegraphics[width=9cm,height=5.5cm]{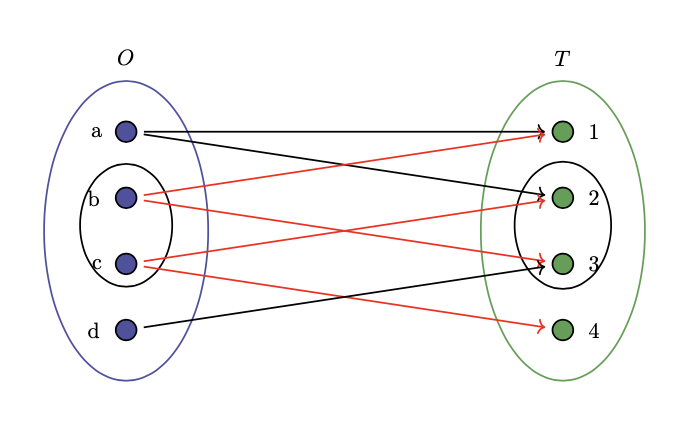}

 %\begin{minipage}{0.45\textwidth} % choose width suitably
\medskip
 %{\footnotesize This is an example of Weighted Directed Network. $O=\{a,b,c,d\}$ is the set of origins, $O_A=\{b,c\}$ is a subset of $O$ with weights $F_{O_A}(b)=0.3$ and  $F_{O_A}(c)=0.6$. $T=\{1,2,3,4\}$ is the set of terminals, $T_B=\{2,3\}$ is a subset of $T$ with weights $G_{T_B}(2)=-0.2$ and $G_{T_B}(3)=0.8$. The subset of edges is $E_L=\{(b,1), (b,3),(c,2),(c,4)\}$, with weights $W_{E_L}=\{0.41, 0.22, -0.15, 0.11\}$ respectively.\par}

 \caption{An example of a weighted directed network}
 \label{Fig1}
%\end{minipage}
\end{figure}

\begin{example}
\label{Exa-digraph}
Fig. \ref{Fig1} is an example of a digraph $G = (O, T, E)$, where $O = \{a, b, c, d\}$ and $T = \{1, 2, 3, 4\}$. The directed arrows represent edges in $G$. So the set of edges $E$ consists of exactly seven edges, say $(a, 1)$, $(a, 2)$, $(b, 1)$, $(b, 3)$, $(c, 2)$, $(c, 4)$, and $(d, 3)$. 

\end{example}

\begin{remark}

For a directed graph $G = (V, E)$, throughout this paper, we denote by $O$ the set of all origins of directed edges in $G$, and by $T$ the set of all terminals of directed edges in $G$. It is clear that $V = O \cup T$. Note that it may occur that there exists a vertex $v$ belonging in $O \cap T$, i.e., $v$ is an origin of a directed edge, and also a terminal of another directed edge. In many places in this paper, we also write $G = (O, T, E)$ to specify the sets of origins and of terminals in $G$.

\end{remark}

We begin by introducing several variants of WDNs. The first variant is a digraph equipped with the set of weights for a subset of origins in the digraph.

\begin{definition}
(Almost origin-weighted directed networks)
\label{Def-AOWDN}

Let $G = (O, T, E)$ be a digraph. $G$ is called an \textit{almost origin-weighted directed network} (written as AOWDN for brevity) if there is a subset $O_A$ of $O$ such that there exists a mapping $F_{O_A} : O_A \to [a, b]$ where $[a, b]$ is a finite interval in the reals $\bR$. 

We use the notation $G = (O, T, E, O_A, F_{O_A})$ to denote the above AOWDN.

\end{definition}

\begin{example}
\label{Exa-AOWDN}
Fig. \ref{Fig1} represents an example of an AOWDN $G = (O, T, E, O_A, F_{O_A})$ in which $O, T, E$ are the same as in Example \ref{Exa-digraph}, $O_A = \{b, c\}$, and $F_{O_A} : O_A \to [0, 1]$ is a mapping of weights of origins in $O_A$ given by $F_{O_A}(b) = 0.3$ and $F_{O_A}(c) = 0.6$. 

\end{example}

\begin{remark}

In real data examples which can be viewed as a digraph $G = (O,T, E)$, most cases assume that there is some weight $F_{O_A}(o)$  associated to each origin $o$ in a subset $O_A$ of $O$, where one can view $O_A$ as a training set. The main aim is to extend the map $F_{O_A}$ to the whole set of origins  $O$, i.e., construct a predictive model for $F_{O_A}$ on $O$ which allows to predict the weight of every vertex in $O$. As an explicit example of an AOWDN, one can take $G$ as a digraph of users and products, where $O$ denotes the set of users, and $T$ denotes the set of products. An edge, say $\fe = (u, p)$ in $G$ is formed if the user $u$ buys the product $p$. One is interested in knowing how fair a user $u$ in $G$ evaluates products in $G$, which we call the weight of $u$. In practice, if the set of users in $G$ evolves over time, and thus at a given time, one only knows a subset of users in $G$, say $O_A$ whose weights are known. Our goal is to predict, in a future time, how fair a user $u$ in the complement of $O_A$ in $O$ is, and thus provides an insight into the dynamic network $G$ of users and products. In this particular example, prediction of weights of users provides an overall evaluation of users towards a fixed set of products in the network $G$.

\end{remark}

Another variant is a digraph equipped with the set of weights for a subset of terminals in the digraph.

\begin{definition}
(Almost terminal-weighted directed networks)

Let $G = (O, T, E)$ be a digraph. $G$ is called an \textit{almost terminal-weighted directed network} (written as ATWDN for short) if there is a subset $T_B$ of $T$ such that there exists a mapping $G_{T_B} : T_B \to [a, b]$ where $[a, b]$ is a finite interval in the reals $\bR$. 

We use the notation $G = (O, T, E, T_B, G_{T_B})$ to denote the above ATWDN. 

\end{definition}

\begin{example}
\label{Exa-ATWDN}

Fig. \ref{Fig1} represents an example of an ATWDN $G = (O, T, E, T_B, G_{T_B})$ in which $O, T, E$ are the same as in Example \ref{Exa-digraph}, $T_B = \{2, 3\}$, and $G_{T_B} : T_B \to [0, 1]$ is a mapping of weights of terminals in $T_B$ given by $G_{T_B}(2) = -0.2$ and $G_{T_B}(4) = 0.8$. 

\end{example}

\begin{remark}

In real data examples which can be viewed as a digraph $G = (O,T, E)$, most cases assume that there is some weight $G_{T_B}(t)$ associated to each terminal $t$ in a subset $T_B$ of $T$, where one can view $T_B$ as a training set. The main aim is to extend the map $G_{T_B}$ to the whole set of terminals $B$, i.e., construct a predictive model for $G_{T_B}$ on $T$.

\end{remark}

The last variant of directed networks considered in this paper is a digraph in which only a subset of edges is equipped with weights.

\begin{definition}
(Almost weighted directed networks)

Let $G = (O,T, E)$ be a digraph. $G$ is called an \textit{almost weighted directed network} (written as AWDN for short) if there is a subset $E_L$ of $E$ such that there exists a mapping $W_{E_L} : E_L \to [a, b]$ where $[a, b]$ is a finite interval in the reals $\bR$. 

We use the notation $G = (O, T, E, E_L, W_{E_L})$ to denote the above AWDN.

\end{definition}

\begin{example}
\label{Exa-AWDN}

Fig. \ref{Fig1} represents an example of an AOWDN $G = (O, T, E, E_L, W_{E_L})$ in which $O, T, E$ are the same as in Example \ref{Exa-digraph}, $E_L = \{(b, 1), (b, 3), (c, 2), (c, 4)\}$ (i.e., all red directed edges in Fig. \ref{Fig1}), and $W_{E_L} : E_L \to [-1, 1]$ is a mapping of weights of edges in $E_L$ given by $W_{E_L}((b, 1)) = 0.41$, $W_{E_L}((b, 3)) = 0.22$, $W_{E_L}((c, 2)) = -0.15$, and $W_{E_L}((c, 4)) = 0.11$. 

\end{example}

\begin{remark}

In real data examples which can be viewed as a digraph $G = (O,T, E)$, most cases assume that there is some invariant $W_{E_L}(\fe)$ associated to each edge $\fe$ in a subset $E_L$ of $E$, where one can view $E_L$ as a training set in edge weight prediction for $G$. The main aim is to extend the map $W_{E_L}$ to the whole set of edges $E$, i.e., construct a predictive model for $W_{E_L}$ on $E$ which allows to predict the weight of every edge in $E$.

\end{remark}

\section{Metrics modulo equivalent relations}

We present in this section a  notion of metric spaces modulo equivalence relations. We first recall a notion of equivalence relations on sets.

\begin{definition}
\label{Def-equivalence-relation}
(Equivalence relation)

Let $X$ be a set. An \textit{equivalence relation}, denoted by $\cong$, on $X$ is a subset of $X \times X$ such that the following are true: 
\begin{itemize}

\item [(i)] (\textbf{Reflexivity}) $(a, a) \in \; \cong$ for every $a \in X$.

\item [(ii)] (\textbf{Symmetry}) $(a, b) \in \; \cong$ if and only if $(b, a) \in \;\cong$.

\item [(iii)] (\textbf{Transitivity}) if $(a, b) \in \; \cong$ and $(b, c) \in \;\cong$ then $(a, c) \in \; \cong$. 

\end{itemize} 

When $(a, b) \in \; \cong$, we say that $a$ is $\cong$--equivalent to $b$. Throughout this paper, in order to signify this relation, we write $a \cong b$ whenever $(a, b) \in \; \cong$. 

\end{definition}

An equivalence relation $\cong$ on a set provides a way to identify \textit{similar} elements in the set. Equivalently if one can find a \textit{measurement} to measure how similar elements in a set are, then one can modify this measurement to introduce an equivalence relation on the set. 

We recall the notion of metrics modulo equivalence relation on a set.

\begin{definition}
\label{def-metric}
(Metric modulo an equivalence relation)

Let $X$ be a set, and $\cong$ an equivalence relation on $X$. A mapping $d : X \times X \to \bR$ is said to be a \emph{metric on $X$ modulo the equivalence relation} $\cong$ if the following condition are satisfied:
\begin{itemize}

\item[(i)] $d(a, b) \ge 0$ for all $a, b \in X$.

\item [(ii)] $d(a, b) = 0$ if and only if $a \cong b$.

\item [(iii)] (\textbf{Symmetry}) $d(a, b) = d(b, a)$ for all $a, b \in X$.

\item [(iv)] (\textbf{Triangle inequality}) for any $a, b ,c \in X$, 
$$d(a, b) \le d(a, c) + d(c, b).$$

\end{itemize}

\end{definition}

A metric modulo an equivalence relation $\cong$ on a set $X$ acts almost like a metric. The only difference between a metric modulo an equivalence relation and a metric on a set is that condition (ii) in Definition \ref{def-metric} is replaced by a stronger condition that $d(a, b) = 0$ if and only if $a = b$. But in studying weight prediction for a given dataset, it is often the case that distinct elements in the dataset share \textit{similar weights}. In this case, it is natural to view that the \textit{distance} between these distinct elements as zero since they are considered to be \textit{equivalent} with respect to the property that the weights associated to them are approximately close. Note that if one uses a usual metric on this dataset, then one cannot identify similarities among distinct elements sharing  almost the same weights. Thus it is more natural to use a metric modulo an equivalence relation on the dataset to study weight predictions for such dataset.

\section{Classes of metrics on variants of WDNs}
\label{Sec-metric}

\subsection{A class of metrics on the set of origins in AOWDNs}
 \label{metric-AOWDN-ss}

In this subsection, we introduce a class of metrics modulo equivalent relations on the set of origins in an AOWDN. This is the first time that such a class of metrics is defined in an AOWDN $G = (O, T, E, O_A, F_{O_A})$. A novel feature of such metrics encodes weight information from the training set $O_A$ to transfer to weight prediction for origins not contained in $O_A$. We begin by introducing a notion of neighbors of origins in an AOWDN. 

\begin{definition} 
\label{def_neighbor_origins}
(neighbors of origins)

Let $G = (O, T, E, O_A, F_{O_A})$ be an AOWDN. Let $o$ be an element in $O$. A neighbor of $o$ is an element $\alpha$ in $O_A$ such that there is a terminal $t \in T$ for which $(o, t)$ and $(\alpha, t)$ both belong to the set of directed edges $E$. 

In notation, let $\mathcal{N}_O(o)$ denote the set of all neighbors of $o$, and set $n_O(o) = \# \mathcal{N}_O(o)$--the number of neighbors of $o$.

\end{definition}

Let $h >0$ be a constant which can be viewed as a tuning parameter. One wants to combine the above notion of neighbors with the constant $h$ to introduce a metric on $O$. Let $o \in O$, and write
\begin{align}
\mathcal{N}_O(o) = \{\alpha_1, \ldots, \alpha_{n_O(o)}\}.
\end{align}

We begin by defining, for each $o \in O$, 
\begin{align}
AvgF(o) = \dfrac{F_{O_A}(\alpha_1) + \cdots + F_{O_A}(\alpha_{n_O(o)})}{n_O(o)}.
\end{align}

For each $o \in O$, let $C_{O,h}(o)$ be the number of neighbors $\alpha$ in $O_A$ of $o$ such that
\begin{align}
|F_{O_A}(\alpha) - AvgF(o)| \le h.
\end{align}

We introduce an equivalence relation on $O$ as follows. We say that $u \cong_O v$ for origins $u, v \in O$ if and only if $C_{O,h}(u) = C_{O,h}(v)$. It is clear that $\cong_O$ is an equivalence relation.

We define a mapping $D_{O, h} : O \times O \to \bR_{\ge 0}$ as follows. For each pair of origins $(u, v) \in O \times O$, define
\begin{align}
D_{O, h}(u, v) =|C_{O,h}(u) - C_{O,h}(v)|.
\end{align}

One obtains the following theorem whose proof will be given in the appendix.

\begin{theorem}
\label{main-thm1}

$D_{O, h}$ is a metric on $O$ modulo the  equivalence relation $\cong_O$.

\end{theorem}

\subsection{A class of metrics on the set of terminals in ATWDNs}
\label{metric-ATWDN-ss}

In this subsection, we introduce a class of metrics modulo equivalent relations on the set of terminals in an ATWDN.

\begin{definition} 
\label{def_neighbor_terminal}
(neighbors of terminals)

Let $G = (O, T, E, T_B, G_{T_B})$ be an ATWDN. Let $t$ be an element in $T$. A neighbor of $t$ is an element $\beta$ in $T_B$ such that there is an origin $o \in O$ for which $(o, t)$ and $(o, \beta)$ both belong to the set of directed edges $E$. 

In notation, let $\mathcal{N}_T(t)$ denote the set of all neighbors of $t$, and set $n_T(t) = \# \mathcal{N}_T(t)$--the number of neighbors of $t$.

\end{definition}

Let $h >0$ be a constant which can be viewed as a tuning parameter. One wants to combine the above notion of neighbors with the constant $h$ to introduce a metric on $T$. Let $t \in T$, and write
\begin{align}
N_T(t) = \{\beta_1, \ldots, \beta_{n_T(t)}\}.
\end{align}

We begin by defining, for each $t \in T$, 
\begin{align}
AvgG(t) = \dfrac{G_{T_B}(\beta_1) + \cdots + G_{T_B}(\beta_{n_T(t)})}{n_T(t)}.
\end{align}

For each $t \in T$, let $C_{T, h}(t)$ be the number of neighbors $\beta$ in $T_B$ of $t$ such that
\begin{align}
|G_{T_B}(\beta) - AvgG(t)| \le h.
\end{align}

We introduce an equivalence relation on $T$ as follows. We say that $p \cong_T q$ for terminals $p, q \in T$ if and only if $C_{T, h}(p) = C_{T, h}(q)$. It is clear that $\cong_T$ is an equivalence relation.

We define a mapping $D_{T, h} : T \times T \to \bR_{\ge 0}$ as follows. For each pair of terminals $(p, q) \in T \times T$, define
\begin{align}
D_{T, h}(p, q) =|C_{T, h}(p) - C_{T, h}(q)|.
\end{align}

One obtains the following result.

\begin{theorem}
\label{main-thm2}

$D_{T, h}$ is a metric on $T$ modulo an equivalence relation $\cong_T$. 

\end{theorem}

The proof of Theorem \ref{main-thm2} will be given in the appendix. 

\subsection{A class of metrics on the set of edges in AWDNs}
\label{metric-AWDN-ss}

In this subsection, we introduce a class of metrics modulo equivalent relations on the set of edges in an AWDN. We begin by introducing a notion of neighbors of edges in an AWDN.

\begin{definition} \label{def_neighbor_edge}
(neighbors of edges)

Let $G = (O, T, E, E_L, W_{E_L})$ be an AWDN. Let $\fe$ be an element in $E$. A neighbor of $\fe$ is an element $\fa$ in $E_L$ such that either $o(\fe) = o(\fa)$ or $t(\fe) = t(\fa)$. 

In notation, let $\mathcal{N}_E(\fe)$ denote the set of all neighbors of $\fe$, and set $n_E(\fe) = \#\mathcal{N}_E(\fe)$--the number of neighbors of $\fe$.

\end{definition}

Let $h >0$ be a constant which can be viewed as a tuning parameter. One wants to combine the above notion of neighbors with the constant $h$ to introduce a metric on $E$. Let $\fe \in E$, and write
\begin{align}
\mathcal{N}_E(\fe) = \{\fa_1, \ldots, \fa_{n_E(\fe)}\}.
\end{align}

We begin by defining, for each $\fe \in E$, 
\begin{align}
AvgW(\fe) = \dfrac{W_{E_L}(\fa_1) + \cdots + W_{E_L}(\fa_{n_E(\fe)})}{n_E(\fe)}.
\end{align}

For each $\fe \in E$, let $C_{E, h}(\fe)$ be the number of neighbors $\fa$ in $E_L$ of $\fe$ such that
\begin{align}
|W_{E_L}(\fa) - AvgW(\fe)| \le h.
\end{align}

We introduce an equivalence relation on $E$ as follows. We say that $\fe \cong_E \fa$ if and only if $C_{E, h}(\fe) = C_{E, h}(\fa)$. It is clear that $\cong_E$ is an equivalence relation.

We define a mapping $D_{E, h} : E \times E \to \bR_{\ge 0}$ as follows. For each pair $(\fe, \fa) \in E \times E$, define
\begin{align}
D_{E, h}(\fe, \fa) =|C_{E, h}(\fe) - C_{E, h}(\fa)|.
\end{align}

We obtain the following theorem whose proof will be given in the appendix.

\begin{theorem}
\label{main-thm3}

$D_{E, h}$ is a metric on $E$ modulo an equivalence relation $\cong_E$. 

\end{theorem}

\section{kNN for variants of WDNs}
\label{Sec-kNN}

In this section, using metrics $D_{O, h}, D_{T, h}$, and $D_{E, h}$, we introduce our modified $k$ nearest neighbors (kNN) method. The method bears a resemblance of the classical kNN for sign prediction. Instead of basing on Euclidean distances as in the classical kNN, we employ our own metrics constructed in Section \ref{Sec-metric}.

\subsection{kNN for AOWDNs}

In this subsection, we introduce a method for predicting a model of $F_{O_A}$ for the whole set $O$ of origin vertices. We use our own metrics constructed in Subsection \ref{metric-AOWDN-ss} for introducing a predictive model for $F_{O_A}$. 

We first take an arbitrary positive integer $k$ of our choice which can be viewed as a tuning parameter for the kNN method proposed here. Let $G = (O, T, E, O_A, F_{O_A})$ be an AOWDN. Our aim is to extend the map $F_{O_A}$ to the whole $O$, i.e., construct the map $F$ on the set of origins such that the restriction of $F$ to $O_A$ is $F_{O_A}$. 

Let $x$ be an arbitrary origin in $O$. Let $d_1, \ldots, d_k$ be the $k$ smallest distance values from $x$ to $O_A$ such that the $d_i$ are nonzero, i.e., the $d_i$ are the $k$ smallest nonzero values among all the values $D_{O, h}(x, \alpha)$ for all $\alpha \in O_A$, where $h$ is a given tuning parameter and $D_{O, h}$ is the metric introduced in Subsection \ref{metric-AOWDN-ss}.

Let $kNN_O(x)$ be the set of all elements $\alpha$ in $O_A$ such that there exists an integer $1 \le i \le k$ for which $d_i = D_{O, h}(x, \alpha)$. We propose to define a predictive model of $F_{O_A}$ as follows. For each $x \in O$, define
\begin{align}
\widehat{F(x)} = \dfrac{\sum_{\alpha \in kNN_O(x)} F_{O_A}(\alpha)}{k}.
\end{align}

\subsection{kNN for ATWDNs}

In this subsection, we introduce a method for predicting a model of $G_{T_B}$ for the set $T$ of terminal vertices. 

Let $k$ be an arbitrary positive integer of our choice which can be viewed as a tuning parameter for the kNN method. Let $G = (O, T, E, T_B, G_{T_B})$ be an ATWDN. Our aim is to extend the map $G_{T_B}$ to the whole set $T$ of terminals, i.e., construct the map $G$ on the set of terminals such that the restriction of $G$ to $T_B$ is $G_{T_B}$. 

Let $t$ be an arbitrary terminal in $T$. Let $d_1, \ldots, d_k$ be the $k$ smallest distance values from $x$ to $T_B$ with respect to the metric $D_{T, h}$ such that the $d_i$ are nonzero, where $h$ is a given tuning parameter. Let $kNN_T(t)$ be the set of elements $\beta$ in $T_B$ such that there exists an integer $1 \le i \le k$ for which $d_i = D_{T, h}(t, \beta)$. We propose to define an extension of $G_{T_B}$ as follows. For each $t \in T$, define
\begin{align}
\widehat{G(t)} = \dfrac{\sum_{\beta \in kNN_T(t)} G_{T_B}(\beta)}{k}.
\end{align}

\subsection{kNN for AWDNs}

In this subsection, we introduce a modified kNN method for predicting a model of $W_{E_L}$ for the set $E$ of directed edges. 

Again we take an arbitrary positive integer $k$ as a tuning parameter for the kNN method. Let $G = (O, T, E, E_L, W_{E_L})$ be an AWDN. Our aim is to extend the map $W_{E_L}$ to the whole set $E$ of edges, i.e., construct the map $W$ on the set of terminals such that the restriction of $W$ to $E_L$ is $W_{E_L}$. 

Let $\fe$ be an arbitrary edge in $E$. Let $d_1, \ldots, d_k$ be the $k$ smallest distance values from $\fe$ to $E_L$ with respect to the metric $D_{E, h}$ such that the $d_i$ are nonzero, where $h$ is a tuning parameter. Let $kNN_E(\fe)$ be the set of elements $\fa$ in $E_L$ such that there exists an integer $1 \le i \le k$ for which $d_i = D_{E, h}(\fe, \fa)$. We propose to define an extension of $W_{E_L}$ as follows. For each $\fe \in E$, define
\begin{align}
\widehat{W(\fe)} = \dfrac{\sum_{\fe \in kNN_E(\fe)} W_{E_L}(\fa)}{k}.
\end{align}

\section{SVM for variants of WDNs}

In this section, we introduce a method for predicting weights in different types of WDNs. The method resembles the classical support-vector machine (SVM) method in regression analysis (see \cite{HTF}) . In order to compute the kernel of SVM model, we introduce a transfer map that embeds the objects we want to study into $\bR$.

Throughout this section, we fix a kernel function $\kappa: \bR \times \bR \to \bR$. There are many choices for such a kernel function such as linear kernel, polynomial kernel, or Gaussian radial basis kernel (see, for example, \cite{HTF}).

\subsection{SVM for AOWDNs}

In this subsection, let $G = (O, T, E, O_A, F_{O_A})$ be an arbitrary AOWDN. Fix a tuning parameter $h > 0$. In order to construct the kernel function for SVM model on $G$, we first define a transfer mapping $\mathcal{T}_O : O \to \bR$ (which allows to view each origin as a real number) of the form
\begin{align}
 \mathcal{T}_O(o)=C_{O,h}(o)
\end{align}
for each $o \in O$, where $C_{O,h}(o)$ is given in Definition \ref{def_neighbor_origins}. 

The SVM model for predicting the weights of origins in $G$ is given by
\begin{align}
\widehat{F(o)}=y(\mathcal{T}_O(o))=\omega_0+\sum_{i=1}^m \omega_i F_{O_A}(\alpha_i) \kappa(\mathcal{T}_O(o), \mathcal{T}_O(\alpha_i)),
\end{align}
where $m$ is the size of the subset $O_A$, $\alpha_1,\ldots, \alpha_m$ are all the origins in $O_A$, and the $w_i$ are coefficients of the SVM model which need to be estimated by using the values $F_{O_A}(\alpha_1), \ldots, F_{O_A}(\alpha_m)$. 

\subsection{SVM for ATWDNs}

In this subsection, let $G = (O, T, E, T_B, G_{T_B})$ be an arbitrary ATWDN. Fix a tuning parameter $h > 0$. The transfer map defined on the set of terminals, denoted as $\mathcal{T}_T : T \to \bR$ (which allows to view each terminal as a real number), is of the form
\begin{align}
  \mathcal{T}_T(t)=C_{T,h}(t)
\end{align}
for each $t \in T$, where $C_{T,h}(t)$ is given in definition \ref{def_neighbor_terminal}.

The SVM model for predicting the weights of terminals in $G$ is given by
\begin{align}
   \widehat{G(t)}=y(\mathcal{T}_T(x))=\omega_0+\sum_{i=1}^n \omega_i G_{T_B}(\beta_i) \kappa(\mathcal{T}_T(t), \mathcal{T}_T(\beta_i)),
\end{align}
where $n$ the size of the subset $T_B$, $\beta_1,\ldots, \beta_n$ are all the terminals in $T_B$, and the $w_i$ are coefficients of the SVM model which need to be estimated by using the values $G_{T_B}(\beta_1), \ldots, G_{T_B}(\beta_n)$.

\subsection{SVM for AWDNs}

In this subsection, let $G = (O, T, E, E_L, W_{E_L})$ be an arbitrary AWDN. Fix a tuning parameter $h > 0$.  The transfer map defined on the set of edges, denoted as $\mathcal{T}_E: E \to \bR$ (which allows to view each edge as a real number), is of the form
\begin{align}
  \mathcal{T}_E(\fe)=C_{E,h}(\fe) 
\end{align}
for each $\fe \in E$, where $C_{E,h}(\fe)$ is given in definition \ref{def_neighbor_edge}.

The SVM model for predicting the weights  of edges in G is given by

\begin{align}
\scriptstyle  \widehat{W(\fe)}=y(\mathcal{T}_E(\fe))=\omega_0+\sum_{j=1}^J \omega_j W_{E_L}(\fa_j) \kappa(\mathcal{T}_E(\fe), \mathcal{T}_E(\fa_j)),
\end{align}
where $J$ is the size of the subset $E_L$, $\fa_1, ..., \fa_J$ are all the edges in $E_L$, and the $w_i$ are coefficients of the SVM model which need to be estimated by using the values $W_{E_L}(\fa_1), \ldots, W_{E_L}(\fa_J)$. 

\section{Experimental analysis on real datasets}

In this section, we apply our modified kNN methods and modified SVM methods to predicting weights of origins, terminals, and edges in three real weighted directed networks--Bitcoin network, Epinions dataset, and WikiSigned dataset. Below we first give a description of each network.
\begin{itemize}

\item \textbf{Bitcoin OTC}. This is a weighted signed directed network of people who trade using Bitcoin on a platform called Bitcoin OTC. The dataset is available at \url{http://snap.stanford.edu/data/soc-sign-bitcoin-otc.html}) (see \cite{KSSF} and \cite{KHMKSF}).

Since users are anonymous, it is necessary to maintain a record of users' reputation to prevent transactions with fraudulent and risky users. Members of Bitcoin OTC rate other members' level of trustfulness on a scale of $-10$ (total distrust) and $+10$ (total trust).

\item \textbf{Epinions}. This dataset was collected by Paolo Massa in a 5-week crawl (November/December 2003) from the Epinions.com Website (see the dataset at \url{http://www.trustlet.orgdownloaded\_epinions.html}) (see \cite{MA}). In Epinions, each user rates the helpfulness of a review on a 1--5 scale,  where $1$ means totally not helpful and $5$ mean totally helpful.

\item \textbf{WikiSigned}. This is a WDN between Wikipedia editors. An edge from an editor $i$ to another editor $j$ represents the degree of trustfulness of $i$ to the edits made by $j$. More details of the dataset could be found in \cite{MCA}. 
\end{itemize}

Note that all datasets above are weighted directed networks in which there is a map of weights $W$ from the set of edges to a closed interval $[a, b]$ in $\bR$. In Bitcoin OTC, the interval is $[-10, 10]$, in Epinions, the interval is $[1, 5]$, and in WikiSigned, the interval is $[-1, 1]$. In the experiments we perform on these datasets, we scale these intervals of edge weights into $[-1, 1]$. In each dataset, we randomly choose $5000$ edges as the set of edges. In order to construct an AWDN from each dataset, we randomly select $3500$ edges out of $5000$ edges, as a training set, say $E_L$ consisting of $3500$ edges equipped with weights. We then use our modified kNN and SVM methods to predict weights of the remaining $1500$ edges. 

We are not aware of any digraph datasets containing weights of origins and terminals. So based on the notion of fairness and goodness introduced in \cite{KSSF}, we associate to each network above weights of origins and terminals. In each network from the above three networks, an edge is represented by $(o, t)$, where $o$ is the origin representing the rater, and $t$ is the terminal of the edge representing the ratee. The weight of origin $o$ we associate here is computed by the \textit{fairness metric} which indicates how fair is the rater $o$ in assessing other terminals. The weight of terminal $t$ we associate here is computed by the \textit{goodness metric} which indicates how good is the ratee $t$ from the viewpoints of other raters. The fairness and goodness metrics are described in more detail in \cite{KSSF}. From the three datasets Bitcoin OTC, Epinions, WikiSigned, we create AOWDNs in which the fairness of origin $o$ represents the weight of $o$, and ATWDNs in which the goodness of terminal $t$ represents the weight of such terminal. In each $AOWDN$, we randomly choose $70\%$ of the set of origins $O$ as the training set $O_A$, and similarly in each ATWDN,  we randomly choose $70\%$ of the set of terminals $T$ as the training set $T_B$. For each $AOWDN$, since the map of weights of origins $F_{O_A}$ is constructed using fairness scores of origins (see Section III(B) in \cite{KSSF} for algorithm and formula to compute the fairness metric), the range of $F_{O_A}$ is $[0, 1]$. For each $ATWDN$, since the map of weights of origins $G_{T_B}$ is constructed using goodness scores of origins (see Section III(B) in \cite{KSSF} for algorithm and formula to compute the goodness metric), the range of $G_{T_B}$ is $[-1, 1]$.

In order to compute the values of metrics $D_{O, h}$, $D_{T, h}$, $D_{E, h}$, we choose the tuning parameter $h$ to be the \textit{standard deviation of all weights} in the training sets $O_A, T_B$, and $E_L$, respectively. To access accuracy in our prediction methods, we use the \textit{mean absolute error} (MAE), and also the root mean square error (RMSE). In Tables \ref{tab1}, \ref{tab2}, \ref{tab3}, and \ref{tab4}, each cell reports a pair of numbers (MAE, RMSE). Our modified kNN and SVM methods perform very well for all predictions. For origin weight prediction, the MAE and RMSE range over the interval $[0, 1]$. For terminal and edge weight predictions, the MAE and RMSE range over the interval $[0, 2]$. We are able to predict origin weights with the MAE ranging in $0.073-0.125$, and the RMSE ranging in $0.138-0.186$. For terminal weight prediction, depending on the networks studied, the MAE ranges in $0.087-0.193$, and the RMSE ranges in $0.163-0.247$. For the edge weight prediction, the MAE ranges in $0.158-0.278$, and the RMSE ranges in $0.312-0.408$, varying for different networks studied.

\begin{table}[htbp]
\caption{Descriptions of Datasets }
\label{tab1}
\begin{center}
\begin{tabular}{|c|c|c|c|c|}
\hline 
\textbf {Network} & \textbf{Origins} & \textbf{Terminals} & \textbf{Edges} & \textbf{\% Positive $E$} \tabularnewline
\hline 
\textbf{Bitcoin OTC} & 1892 & 2070 & 5000 & 90.26\%\tabularnewline
\hline 
\textbf{Epinions} & 2860 & 4059 & 5000 & 86.20\%\tabularnewline
\hline 
\textbf{WikiSigned} & 4704 & 4296 & 5000 & 95.08 \% \tabularnewline
\hline 
\end{tabular}
\label{tab1}
\end{center}
\end{table}

\begin{table}[htbp]
\begin{centering}
\caption{Results of Predicting Weights of Origins
\label{tab2}}
\begin{tabular}{|c|c|c|c|c|c|c|}
\hline 
 \textbf{Network}& \textbf{kNN} & \textbf{SVM}\tabularnewline
\hline 
\textbf{Bitcoin OTC} & (0.075, 0.139) & (0.073, 0.138) \tabularnewline
\hline 
\textbf{Epinions} & (0.125, 0.163) & (0.116, 0.186) \tabularnewline
\hline 
\textbf{WikiSigned} & (0.095, 0.155) & (0.081, 0.158) \tabularnewline
\hline 
\end{tabular}
\par\end{centering}
\end{table}

\begin{table}[htbp]
\begin{centering}
\caption{Results of Predicting Weights of Terminals
\label{tab3}}
\begin{tabular}{|c|c|c|c|c|c|c|}
\hline 
 \textbf{Network} & \textbf{kNN} & \textbf{SVM} \tabularnewline
\hline 
\textbf{Bitcoin OTC} & (0.099, 0.163) & (0.087, 0.163) \tabularnewline
\hline 
\textbf{Epinions} & (0.193, 0.221) & (0.183 0.247) \tabularnewline
\hline 
\textbf{WikiSigned} & (0.134, 0.209) & (0.096, 0.218)  \tabularnewline
\hline 
\end{tabular}
\par\end{centering}
\end{table}

\begin{table}[htbp]
\begin{centering}
\caption{Results of Predicting Edge Weights. 
\label{tab4}}
\begin{tabular}{|c|c|c|c|c|c|c|}
\hline 
\textbf{Network}& \textbf{kNN} & \textbf{SVM} \tabularnewline
\hline 
\textbf{Bitcoin OTC} & (0.193, 0.312) & (0.158, 0.315) \tabularnewline
\hline 
\textbf{Epinions} & (0.278, 0.353) & (0.245, 0.408)\tabularnewline
\hline 
\textbf{WikiSigned} & (0.189, 0.312) & (0.158, 0.315)  \tabularnewline
\hline 
\end{tabular}
\par\end{centering}
\end{table}

\section{Conclusions}

Our paper proposes novel geometric approaches to predict edge weights in weighted directed networks. Our paper also studies weight prediction for vertices in networks, which has not been investigated before, to the best of our knowledge.

Our main contributions are as follows:
\begin{itemize}

\item \textbf{Variants of weighted directed networks}

Our work is the first work that has introduced several variants of directed networks equipped with weights of origins, terminals, or edges. We call these networks \textit{almost-weighted origin directed networks}, \textit{almost-weighted origin directed networks}, \textit{almost-weighted edge directed networks}, respectively. These types of networks are very suitable for modeling real-world datasets since in most cases, one only knows weights of certain subsets of the set of origins, terminals, or edges, especially for \textit{temporal or dynamic networks} in which the graph structures change over time. For the purpose of predicting weights, it is very useful to have weight information of the network at a given time, which can be viewed as a training set of weights.

\item \textbf{Novel geometric approaches}

We introduce a metric geometry approach to studying weight prediction problems in digraphs. We introduce several classes of metrics modulo equivalent relations on different types of weighted digraphs.

\item \textbf{Modified kNNs and SVMs}

We introduce modified $k$ nearest neighbors method and support-vector machine methods for predicting weights in digraphs. These methods base on the metric geometric structures of digraphs that we introduce in this work.

\end{itemize}

\section{Appendix}

In this appendix, we prove Theorems \ref{main-thm1}, \ref{main-thm2}, and \ref{main-thm3}.

\begin{proof}[A. Proof of Theorem \ref{main-thm1}]

It is clear that $D_{O, h}(u, v) \ge 0$ for any origins $u, v \in O$. Thus (i) in Definition \ref{def-metric} follows. Condition (iii) in Definition \ref{def-metric} is straightforward. By definition of $\cong_{O}$, $C_{O, h}(u) = C_{O, h}(v)$ if and only if $u \cong_O v$. Thus $D_{O, h}(u, v) = |C_{O, h}(u) - C_{O, h}(v)| = 0$. Thus 
if and only if $u \cong_O v$, which proves (ii) in Definition \ref{def-metric}. 

For any origins $u, v, w \in O$, we see that
\begin{align*}
D_{O, h}(u, w) &= |C_{O, h}(u) - C_{O, h}(w)| \\
&=  |(C_{O, h}(u) - C_{O, h}(v)) + (C_{O, h}(v) - C_{O, h}(w))|  \\
&\le   |(C_{O, h}(u) - C_{O, h}(v))| + |C_{O, h}(v) - C_{O, h}(w)| \\
&= D_{O, h}(u, v) + D_{O, h}(v, w),
\end{align*}
which verifies (iv) in Definition \ref{def-metric}. Thus $D_{O, h}$ is a metric on $O$ modulo the equivalence relation $\cong_O$.

\end{proof}

\begin{proof}[B. Proof of Theorem \ref{main-thm2}]

It is clear that $D_{T, h}(p, q) \ge 0$ for any terminals $p, q \in T$. Thus (i) in Definition \ref{def-metric} follows. Condition (iii) in Definition \ref{def-metric} is straightforward. By definition of $\cong_{T}$, $C_{T, h}(p) = C_{T, h}(q)$ if and only if $p \cong_T q$. Thus $D_{T, h}(p, q) = |C_{T, h}(p) - C_{T, h}(q)| = 0$. Thus 
if and only if $p \cong_T q$, which proves (ii) in Definition \ref{def-metric}. 

For any terminals $p, q, r \in O$, we see that
\begin{align*}
D_{T, h}(p, r) &= |C_{T, h}(p) - C_{T, h}(r)| \\
&=  |(C_{T, h}(p) - C_{T, h}(q)) + (C_{T, h}(q) - C_{T, h}(r))|  \\
&\le   |(C_{T, h}(p) - C_{T, h}(q))| + |C_{T, h}(q) - C_{T, h}(r)| \\
&= D_{T, h}(p, q) + D_{T, h}(q, r),
\end{align*}
which verifies (iv) in Definition \ref{def-metric}. Thus $D_{T, h}$ is a metric on $T$ modulo the equivalence relation $\cong_T$.

\end{proof}

\begin{proof}[B. Proof of Theorem \ref{main-thm3}]

It is clear that $D_{E, h}(\fa, \fe) \ge 0$ for any edges $\fa, \fe \in T$. Thus (i) in Definition \ref{def-metric} follows. Condition (iii) in Definition \ref{def-metric} is straightforward. By definition of $\cong_{E}$, $C_{E, h}(\fa) = C_{E, h}(\fe)$ if and only if $\fa \cong_E \fe$. Thus $D_{E, h}(\fa, \fe) = |C_{E, h}(\fa) - C_{E, h}(\fe)| = 0$. Thus 
if and only if $\fa \cong_E \fe$, which proves (ii) in Definition \ref{def-metric}. 

For any terminals $\fa, \fb, \fe \in O$, we see that
\begin{align*}
D_{E, h}(\fa, \fb) &= |C_{E, h}(\fa) - C_{E, h}(\fb)| \\
&=  |(C_{E, h}(\fa) - C_{E, h}(\fe)) + (C_{E, h}(\fe) - C_{E, h}(\fb))|  \\
&\le   |(C_{E, h}(\fa) - C_{E, h}(\fe))| + |C_{E, h}(\fe) - C_{E, h}(\fb)| \\
&= D_{E, h}(\fa, \fe) + D_{E, h}(\fe, \fb),
\end{align*}
which verifies (iv) in Definition \ref{def-metric}. Thus $D_{E, h}$ is a metric on $E$ modulo the equivalence relation $\cong_E$.

\end{proof}

%\section*{Acknowledgements}

%DQNN was partially supported by DARPA grant N66001-17-1-4041.

\bibliographystyle{ieeetr}
\bibliography{reference}

\begin{thebibliography}{10}

\bibitem{TB}
V.~A. Traag and J.~Bruggeman, ``Community detection in networks with positive
  and negative links,'' {\em Physical Review E}, vol.~80, no.~3, p.~036115,
  2009.

\bibitem{YG}
B.~Yan and S.~Gregory, ``Detecting community structure in networks using edge
  prediction methods,'' {\em Journal of Statistical Mechanics: Theory and
  Experiment}, vol.~2012, no.~09, p.~P09008, 2012.

\bibitem{KSS}
S.~Kumar, F.~Spezzano, and V.~Subrahmanian, ``Accurately detecting trolls in
  slashdot zoo via decluttering,'' in {\em 2014 IEEE/ACM International
  Conference on Advances in Social Networks Analysis and Mining (ASONAM 2014)},
  pp.~188--195, IEEE, 2014.

\bibitem{KHMKSF}
S.~Kumar, B.~Hooi, D.~Makhija, M.~Kumar, C.~Faloutsos, and V.~Subrahmanian,
  ``Rev2: Fraudulent user prediction in rating platforms,'' in {\em Proceedings
  of the Eleventh ACM International Conference on Web Search and Data Mining},
  pp.~333--341, 2018.

\bibitem{BRMA}
E.~Bakshy, I.~Rosenn, C.~Marlow, and L.~Adamic, ``The role of social networks
  in information diffusion,'' in {\em Proceedings of the 21st international
  conference on World Wide Web}, pp.~519--528, 2012.

\bibitem{SJ}
M.~Shafaei and M.~Jalili, ``Community structure and information cascade in
  signed networks,'' {\em New Generation Computing}, vol.~32, no.~3-4,
  pp.~257--269, 2014.

\bibitem{RSH}
A.~Roy, C.~Sarkar, J.~Srivastava, and J.~Huh, ``Trustingness \&
  trustworthiness: A pair of complementary trust measures in a social
  network,'' in {\em 2016 IEEE/ACM International Conference on Advances in
  Social Networks Analysis and Mining (ASONAM)}, pp.~549--554, IEEE, 2016.

\bibitem{WAS}
Z.~Wu, C.~C. Aggarwal, and J.~Sun, ``The troll-trust model for ranking in
  signed networks,'' in {\em Proceedings of the Ninth ACM international
  conference on Web Search and Data Mining}, pp.~447--456, 2016.

\bibitem{LHK}
J.~Leskovec, D.~Huttenlocher, and J.~Kleinberg, ``Predicting positive and
  negative links in online social networks,'' in {\em Proceedings of the 19th
  international conference on World wide web}, pp.~641--650, 2010.

\bibitem{KSSF}
S.~Kumar, F.~Spezzano, V.~Subrahmanian, and C.~Faloutsos, ``Edge weight
  prediction in weighted signed networks,'' in {\em 2016 IEEE 16th
  International Conference on Data Mining (ICDM)}, pp.~221--230, IEEE, 2016.

\bibitem{HTF}
T.~Hastie, R.~Tibshirani, and J.~Friedman, {\em The elements of statistical
  learning: data mining, inference, and prediction}.
\newblock Springer Science \& Business Media, 2009.

\bibitem{MA}
P.~Massa and P.~Avesani, ``Trust-aware recommender systems,'' in {\em
  Proceedings of the 2007 ACM conference on Recommender systems}, pp.~17--24,
  2007.

\bibitem{MCA}
S.~Maniu, B.~Cautis, and T.~Abdessalem, ``Building a signed network from
  interactions in wikipedia,'' in {\em Databases and Social Networks},
  pp.~19--24, 2011.

\end{thebibliography}

\end{document}